\documentclass{PoS}
\usepackage{slashed}
\usepackage{amsmath}

\title{QFT with FDR}

\ShortTitle{QFT with FDR}

\author{\speaker{Roberto Pittau}\\
        Departamento de F\'isica Te\'orica y del Cosmos and CAFPE, Universidad de Granada, Campus Fuentenueva s.n., E-18071
Granada, Spain\\
        E-mail: \email{pittau@ugr.es}}

\abstract{I discuss the latest developments of FDR in the context of quantum field theory calculations relevant for high-energy particle physics phenomenology. In particular, I focus on NNLO computations and on the use of FDR in connection with effective field theories.}

\FullConference{14th International Symposium on Radiative Corrections (RADCOR2019)\\ 
9-13 September 2019\\
		Palais des Papes, Avignon, France}

\begin{document}
\def\d#1{D_{#1}}
\def\db#1{\bar D_{#1}}
\def\tld#1{\tilde {#1}}
\newcommand{\aem}{\alpha_{\scriptscriptstyle E\!M}}
\newcommand{\qbarslash}{\,\overline{\slashed{q}}}
\newcommand{\I}{\,\mathfrak{I}}
\newcommand{\D}{\,{D}}
\newcommand{\Dbar}{\bar D}
\newcommand{\qbar}{\bar q}
\newcommand{\amp}{\,\mathcal{M}}
\newcommand{\mur}{\mu_{\scriptscriptstyle R}}
\newcommand{\nl}{\nonumber \\}
\newcommand{\bfig}{\begin{center}\begin{picture}}
\newcommand{\efig}[1]{\end{picture}\\{\small #1}\end{center}}
\newcommand{\flin}[2]{\ArrowLine(#1)(#2)}
\newcommand{\ghlin}[2]{\DashArrowLine(#1)(#2){5}}
\newcommand{\wlin}[2]{\DashLine(#1)(#2){2.5}}
\newcommand{\zlin}[2]{\DashLine(#1)(#2){5}}
\newcommand{\glin}[3]{\Photon(#1)(#2){2}{#3}}
\newcommand{\gluon}[3]{\Gluon(#1)(#2){5}{#3}}
\newcommand{\lin}[2]{\Line(#1)(#2)}
\newcommand{\sof}{\SetOffset}
\newcommand{\bqa}{\begin{eqnarray}}
\newcommand{\eqa}{\end{eqnarray}}
\newcommand{\nn}{\nonumber}
\newcommand{\als}{\alpha_S}
\newcommand{\dbar}{\bar D}
\newcommand{\DRFDR}{{\rm D_{\rm {\scriptscriptstyle R}eg}}^{\rm\!\!\!\!\!\!\!\! \scriptscriptstyle FDR}}
\newcommand\toGP{\,~{\to}^{\vphantom{a}^{\hskip -11pt {\mbox{\tiny \rm GP}}}}~\,}
\newcommand\toSIC{\,~{\to}^{\vphantom{a}^{\hskip -11pt {\mbox{\tiny \rm SIC}}}}~\,}
\newcommand\toGS{\,~{\to}^{\vphantom{a}^{\hskip -11pt {\mbox{\tiny \rm GS}}}}~\,}
\newcommand\qot{q_{1\!2}}
\newcommand\qots{q^2_{1\!2}}
\newcommand\qotsb{\bar q^2_{1\!2}}
\newcommand{\Lr}{L_{\scriptscriptstyle R}}
\newcommand{\mj}{{\{m_j\}}}
\newcommand{\mjone}{{\{m_1\}}}
\newcommand{\mjtwo}{{\{m_2\}}}

\section{Introduction}
The overwhelming need for precise theoretical predictions in collider phenomenology requires a drastic rethinking of the current methodologies employed in perturbative quantum field theory (pQFT) calculations. In particular, the computational complexity grows very fast as a function of the perturbative order. The main reason for such a challenge is the appearance of intermediate expressions that diverge in the limit of hard/soft/collinear kinematic configurations, while the final result is free of singularities. The customary way to deal with such a problem is based on analytically continuing to $d$ dimensions the divergent loop and phase-space integrals, while taking the $d\to 4$ limit at the end of the calculation \cite{Bollini:1972ui,tHooft:1972tcz}. Unfortunately, such integrals do not lead themselves to a direct numerical computation and a huge amount of work is usually required to analytically extract the singularities.
Recently, fully four-dimensional methods have been introduced to overcome these
complications \cite{Cherchiglia:2010yd,Pittau:2012zd,Sborlini:2016gbr,Gnendiger:2017pys}. So far, they have been implemented and tested up to NNLO for certain classes of processes \cite{Cherchiglia:2012zp},
\cite{Donati:2013iya,Pittau:2013qla,Donati:2013voa,Page:2015zca,Page:2018ljf},
\cite{Driencourt-Mangin:2017gop,Driencourt-Mangin:2019aix}.
In this contribution I review the present status of FDR \cite{Pittau:2012zd} in high-energy pQFT calculations as well as in the context of nonrenormalizable effective field theories (EFT).

\section{Using FDR in pQFT calculations}
\begin{figure}[t]
\vspace{-9cm}
\hskip -18pt
\includegraphics[width=1.9\textwidth, angle= {0}]{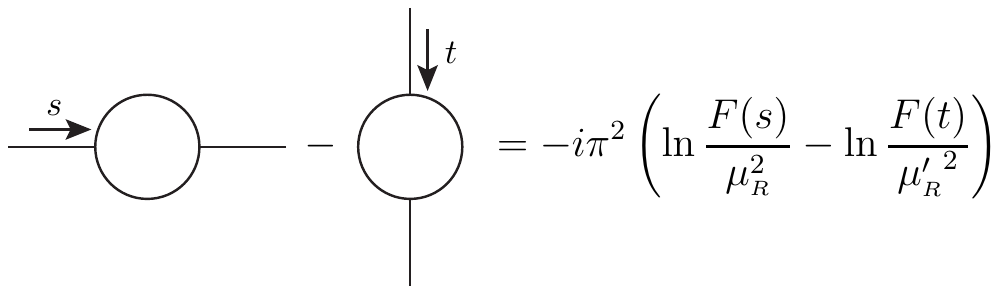}
\vspace{-9cm}
\caption{Combining two one-loop scalar functions depending on different kinematic invariants $s$ and $t$ gives the correct UV finite result $-i \pi^2 \left(\ln \frac{F(s)}{F(t)} \right)$ only if $\mur^\prime= \mur$. Since it is hard to believe that the $s$-channel diagram knows something about the $t$-channel one (and vice versa), one further assumption is $\mur^\prime= \mur= {\rm constant}$.}
\label{fig:1}
\end{figure}

The ultraviolet (UV) problem is solved in FDR by subtracting UV divergences directly at the integrand level. This is achieved by introducing a suitable linear integral operator, denoted by $\int [d^4q]$, whose action on a UV divergent integrand produces a finite result, which only depends upon the renormalization scale $\mur$. For instance
\bqa
\label{eq:I1}
I^1_{\scriptscriptstyle \rm FDR} := \int [d^4q] \frac{1}{(\qbar^2-M^2)^2}= -i \pi^2 \ln\frac{M^2}{ \mur^2}.
\eqa
The notation used in \eqref{eq:I1} is ${\qbar^2:= q^2-\mu^2}$, where $\mu^2$ is an auxiliary mass needed to extract the UV divergent piece by partial fractioning,
\bqa
\frac{1}{(\qbar^2-M^2)^2}= {\left[\frac{1}{\qbar^4} \right]}
+\frac{M^2}{\qbar^2(\qbar^2-M^2)^2}+\frac{M^2}{\qbar^4(\qbar^2-M^2)}.
\eqa
The term between square brackets is UV divergent and only depends on $\mu$. For this reason it is considered unphysical and it is annihilated by $\int [d^4q]$.
Subsequently, after taking the asymptotic $\mu \to 0$ limit, $\mu$ is identified with $\mur$,
\bqa
\int[d^4q] \frac{1}{(\qbar^2-M^2)^2} :=
\left.{\lim_{\mu \to 0}} \int d^4q  
\left(\frac{M^2}{\qbar^2(\qbar^2-M^2)^2}+\frac{M^2}{{ \qbar^4}(\qbar^2-M^2)}\right)\right|_{\mu = \mur} \!\!\!= -i \pi^2 \ln\frac{M^2}{ \mur^2}.
\eqa
Note that $\mur$ is not a cut-off. For example, it appears logarithmically also in quadratically divergent integrals,
\bqa
\int [d^4q] \frac{1}{(\qbar^2-M^2)}= -i \pi^2 M^2 \left(\ln\frac{M^2}{ \mur^2}-1\right).
\eqa
More in general, the structure of a UV divergent $\ell$-loop FDR integral is a polynomial of degree $\ell$ in $\ln (\mur^2)$, 
\bqa
\label{eq:Iell}
I^\ell_{\scriptscriptstyle \rm FDR}= \sum_{k=0}^\ell c_k \Lr^k,\hskip 10pt \Lr:= \ln(\mur^2). 
\eqa
For instance \cite{Donati:2013voa}
\bqa
\label{eq:I2}
&&I^2_{\scriptscriptstyle \rm FDR} :=\int[d^4q_1] [d^4q_2] \frac{1}{(\qbar_1^2-m_1^2)^2 (\qbar_2^2-m_2^2)
  ((q_1+q_2)^2-\mu^2)}= \nl
&&\hskip 40pt \pi^4\left\{
\frac{i}{\sqrt{3}} \left[ {\rm Li}_2\left(e^{i \frac{\pi}{3}}\right)-{\rm Li}_2\left(e^{-i \frac{\pi}{3}}\right) \right]
- {\rm Li}_2\left(1-\frac{m_2^2}{m_1^2} \right)
- \frac{1}{2} \ln^2 \frac{\mur^2}{m_1^2}- \ln \frac{\mur^2}{m_1^2}
\right\}.
\eqa
It is important to realize that internal consistency requires $\mur$ to be independent of kinematics and identical in all loop functions. This guarantees correct cancellations when combining integrals, as illustrated in Fig. \ref{fig:1}.

FDR is compatible with pQFT calculations because it respects gauge invariance and unitarity. In fact, FDR integrals are shift invariant objects which maintain the cancellations between reconstructed denominators and propagators required to prove graphical Ward-identities \cite{Donati:2013iya,Donati:2013voa}.  
In addition, unitarity is enforced by a special treatment of the Lorentz indices external to the divergent sub-diagrams.\footnote{See appendix A of \cite{Page:2018ljf}.} This guarantees that $\ell$-loop structures give the same result also when embedded in $(\ell+1)$-loop calculations.

\begin{figure}[t]
\vspace{-8.5cm}
\hskip -118pt
\includegraphics[width=1.9\textwidth, angle= {0}]{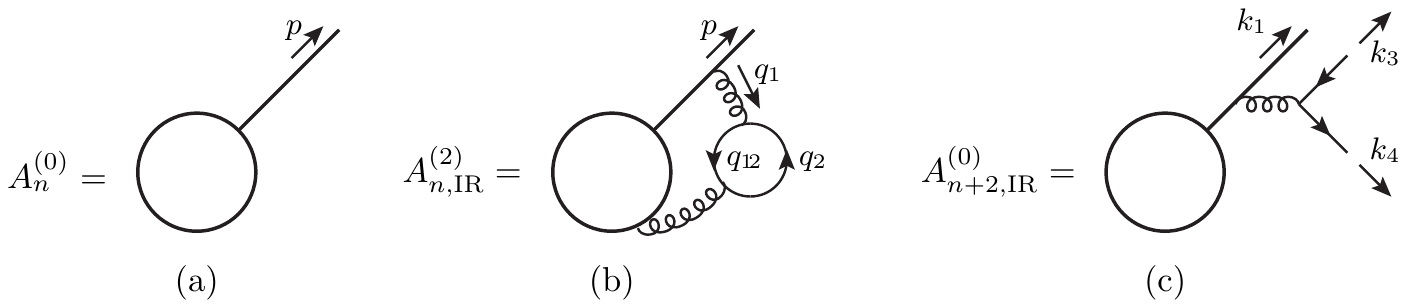}
\vspace{-9.5cm}
\caption{The lowest order amplitude (a), the IR divergent final-state virtual quark-pair correction (b) and the IR divergent real component (c).
The blob stands for the emission of $n\!-\!1$ particles.
Additional IR finite corrections are created if the gluon which splits into $q \bar q$ is emitted by the blob.}
\label{fig:2}
\end{figure}

In the presence of infrared (IR) divergences, the insertion of $\mu^2$ in the denominators acts as a regulator of the IR behavior of the loop integrals. Consistency requires the addition of a mass $\mu$ also in the phase-space integration of the real counterpart. Things can be arranged in such a way that the real and virtual parts combine to give the correct result. This has been explicitly shown in \cite{Page:2018ljf}, where the fully inclusive NNLO final-state quark-pair corrections depicted in Fig. \ref{fig:2} have been computed without relying, explicitly or implicitly, on dimensional regularization.
The physical observable is
\bqa
{\sigma^{\mbox{\tiny NNLO}}} = \sigma_{B}+\sigma_{V}+\sigma_{R},
\eqa
with
\bqa
\label{eq:sigmaBVR}
\sigma_B &\propto& \int d \Phi_n\, \sum_{\rm spin}
|A^{(0)}_{n}|^2, \nl
\sigma_V &\propto& \int d \Phi_n\,
\sum_{\rm spin}
\left\{
 A^{(2)}_{n} (A^{(0)}_{n})^\ast 
+A^{(0)}_{n} (A^{(2)}_{n})^\ast 
\right\}, \nl
\sigma_R &\propto& \int d \Phi_{n+2}\,
\sum_{\rm spin}
\left\{
A^{(0)}_{n+2} (A^{(0)}_{n+2})^\ast
\right\},
\eqa
and the known  $H \to b \bar b + jets$ and  $\gamma^\ast \to jets$ results 
\bqa
\hskip -16pt 
\Gamma_{\mbox{\tiny\it H} \to b \bar b+jets}^{\mbox{\tiny NNLO}}(y_b) &=&  \Gamma_2^{(0)}(y_b)
\left\{\!1+
(\alpha_S/{4\pi})^2 C_F N_F 
\left(
2\, { \ln^2 \frac{m_b^2}{M^2_H}}
-\frac{26}{3} { \ln \frac{m_b^2}{M^2_H}}
+ 8 \zeta_3 + 2 \pi^2 -\frac{62}{3}
\right)\!
\right\},  \nl
\sigma^{\mbox{\tiny NNLO}}_{e^+ e^- \to jets}
&=& \sigma_2^{(0)}
\left\{1+
 (\alpha_S/{4\pi})^2 C_F N_F 
\left(
8 \zeta_3 -11
\right)
\right\} 
\eqa
are reproduced by FDR when adding the three pieces. 

One among the most important consequences of an integrand-level subtraction such as in \eqref{eq:I1} and \eqref{eq:I2} is that an order-by-order removal of the UV infinities is avoided. To illustrate this, consider a renormalizable Lagrangian ${\cal L}^{\scriptstyle \rm R}(p_1,\ldots,p_m)$. It becomes predictive only after fixing the bare parameters $p_k$ in terms of experimental observables ${\cal O}_k^{\rm EXP}$ computed at the loop level $\ell$ one is working,
\bqa
\tilde p_k(\mur):= p_k^{\rm TH,\ell-loop}({\cal O}_1^{\rm EXP},\ldots,{\cal O}_m^{\rm EXP}\!,\mur),~~~ k= 1 \div m.
\eqa
At this stage $\mur$ still appears, but it cancels out when computing a further independent observable ${\cal O}^{\rm TH,\ell-loop}_{m+1}$,
\bqa
\label{eq:reng}
\frac{d {\cal O}^{\rm TH,\ell-loop}_{m+1}\big(\tilde p_1(\mur),\ldots,\tilde p_m(\mur),\mur\big) }{d \mur} =   0.
\eqa
Eq. \eqref{eq:reng} is nothing but the renormalization group equation for 
${\cal O}^{\rm TH,\ell-loop}_{m+1}$, that is obtained right away in FDR with no need of constructing a counterterm Lagrangian $\Delta{\cal L}^{\scriptstyle\, \rm Counterterms}$ such that
\bqa
{\cal L}^{\scriptstyle \rm R}(p_1,\ldots,p_m)= {\cal L}^{\scriptstyle\, \rm Renormalized}+\Delta {\cal L}^{\scriptstyle\, \rm Counterterms}.
\eqa

\section{Using FDR in EFT}
In a pQFT described by a nonrenormalizable Lagrangian
${\cal L}^{\scriptstyle \rm N}(p_1,\ldots,p_m)$ one has
\bqa
\label{eq:noreng}
\frac{d {\cal O}^{\rm TH,\ell-loop}_{m+1}\big(\tilde p_1(\mur),\ldots,\tilde p_m(\mur),\mur\big) }{d \mur} \ne   0.
\eqa
However, if one could infer $\mur$ by other means,
${\cal L}^{\scriptstyle \rm N}(p_1,\ldots,p_m)$ would describe as it stands a legitimate EFT. This means that introducing higher-dimensional operators $\Delta {\cal L}_{\scriptstyle \rm  HDO}$ to reabsorb the UV infinities generated by the loop expansion of the interactions contained in  ${\cal L}^{\scriptstyle \rm N}(p_1,\ldots,p_m)$,
\bqa
\label{eq:LtoLHDO}
{\cal L}^{\scriptstyle \rm N}(p_1,\ldots,p_m) \to {\cal L}^{\scriptstyle \rm N}(p_1,\ldots,p_m)+\Delta {\cal L}_{\scriptstyle \rm  HDO},
\eqa
is not necessarily needed in FDR \cite{Pittau:2013ica}.
In the following, I describe the order-by-order conditions under which a value of $\mur$ can be found that matches a given renormalizable model onto ${\cal L}^{\scriptstyle \rm N}(p_1,\ldots,p_m)$ without the addition of $\Delta {\cal L}_{\scriptstyle \rm  HDO}$ \cite{Pittau:2019bba}.

\begin{figure}[t]
\vspace{-6.8cm}
\hskip -58pt
\includegraphics[width=1.9\textwidth, angle= {0}]{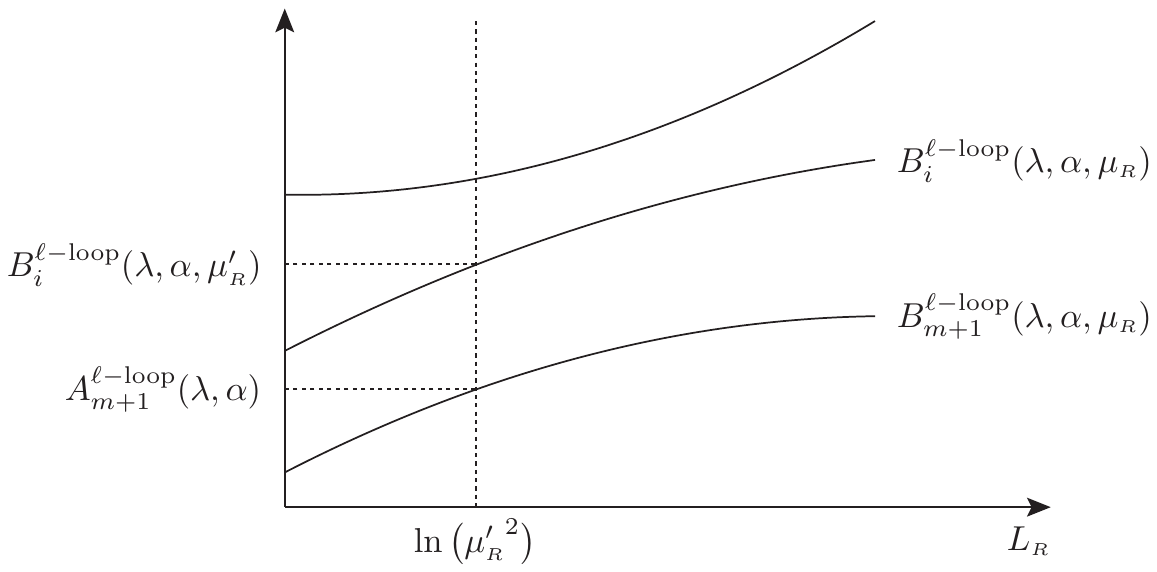}
\vspace{-8.5cm}
\caption{The space of the predictions of ${\cal L}^{\scriptstyle \rm N}(p_1,\ldots,p_m)$ as a function of $\Lr$ in \eqref{eq:Iell}.  The all-order expansions of the
  amplitudes labelled by the subscript $m+1$ is given in \eqref{eq:eqAB}. The index $i> m+1$ denotes an observable different from those employed to determine the Lagrangian's parameter and $\mur^\prime$.}
\label{fig:3}
\end{figure}

The starting point for the matching is the equation
\bqa
\label{eq:init0}
B_{m+1}({0},\alpha,\mur) = A_{m+1}({0},\alpha),
\eqa
where $B_{m+1}({0},\alpha,\mur)$ and $A_{m+1}({0},\alpha)$ are the amplitudes
for the observable ${\cal O}_{m+1}$ in \eqref{eq:noreng} and \eqref{eq:reng} computed in the effective and renormalizable models, respectively, and $\alpha$ is a coupling constant. I assume that ${\cal O}_{m+1}$ refer to a zero-energy measurement and that a ${\cal L}^{\scriptstyle \rm N}(p_1,\ldots,p_m)$ exist such that \eqref{eq:init0} holds.
If $N$ is the number of independent kinematic invariants $s_n$, the energy dependence of ${\cal O}_{m+1}$ can be described in terms of the ratios
\bqa
\lambda_n=s_n/{M^2_n},~~~n=1 \div N,
\eqa
where the $M_n$ are mass scales parameterizing the range of validity of the effective description. The aim is to find an order-by-order solution $\mur= \mur^\prime$ for which \eqref{eq:init0} persist also when $\lambda := \{\lambda_1,\ldots,\lambda_N\} \ne 0$,
\bqa
\label{eq:match}
B^{\rm \ell-loop}_{m+1}(\lambda,\alpha,\mur^\prime) = A^{\rm \ell-loop}_{m+1}(\lambda,\alpha),
\eqa
as illustrated schematically in Fig. \ref{fig:3}.
To determine $\mur^\prime$ one writes down the all-order expansions of
$A_{m+1}$ and $B_{m+1}$,
\begin{subequations}\label{eq:eqAB}
\bqa
\label{eq:eqABa}
A_{m+1}(\lambda,\alpha) &=&
K(\alpha)
+K(\alpha) \sum_{j=1}^{\infty} A_{0j}^\mj \lambda^\mj  
 + K(\alpha)\!\!\sum_{i,j=1}^{\infty} A_{ij}^\mj \alpha^i \lambda^\mj, \\
\label{eq:eqABb}
B_{m+1}(\lambda,\alpha,\mur)&=&
K(\alpha) + K(\alpha)\!\!\sum_{\substack{i,j=1\\0 \leq k \leq i}}^{\infty}\! B_{ijk}^\mj \alpha^i \lambda^\mj \Lr^k, 
\eqa
\end{subequations}
where $K(\alpha)$ implements the initial condition \eqref{eq:init0},
\bqa
\label{eq:lemat}
B_{m+1}(0,\alpha,\mur) = A_{m+1}(0,\alpha)= K(\alpha).
\eqa
$A_{0j}^\mj$, $A_{ij}^\mj$, $B_{ijk}^\mj$ are perturbative coefficients,
in which $i$ refers to the $\alpha$ expansion, whereas $j$
denotes the power degree of the products of $\lambda_n$ multiplying the coefficients.
The notation $$\mj:=(m_{j1},m_{j2},\ldots,m_{jN})$$ symbolizes an assignment of  integer numbers $m_{jn} \ge 0 $ fulfilling
$
\sum_{n=1}^N m_{jn}= j
$,
and a sum over all possible assignments is understood when contracting with
$
\lambda^\mj := \prod_{n=1}^N \lambda_n^{m_{jn}}
$.
For instance, if $N=2$,
$
  A_{02}^\mjtwo \lambda^\mjtwo =
  A_{02}^{(2,0)} \lambda_1^2
 +A_{02}^{(0,2)} \lambda_2^2
 +A_{02}^{(1,1)} \lambda_1 \lambda_2
$. 
The coefficients in \eqref{eq:eqAB} may involve functions of $s_n$ singular at $\lambda= 0$, such as $\ln s_n$ or ${s_n}^{-\frac{1}{2}}$, but \eqref{eq:lemat} requires
\bqa
A_{0j}^\mj \lambda^\mj \to 0,~~~
A_{ij}^\mj \lambda^\mj \to 0,~~~
B_{ijk}^\mj \lambda^\mj \to 0 
\eqa
when $\lambda \to 0$.
Furthermore, $B_{m+1}$ in \eqref{eq:eqABb} depends on $\lambda$ only through loop corrections, unlike $A_{m+1}$. Typically, the second term in the r.h.s. of \eqref{eq:eqABa} is generated by Taylor expanding the tree-level propagators $1/(s_n-M^2_n)$ of the exact theory, that are absent in the effective model. Note also that the dependence upon $\mur$ is driven by \eqref{eq:Iell}.
Solutions to \eqref{eq:match} are found by replacing its two sides by
\eqref{eq:eqABa} computed with $(i \le \ell, j \le \ell)$ and
\eqref{eq:eqABb} truncated at $(i \le \ell+1,j \le \ell, k \ge i-\ell)$,
and allowing $\Lr$ in \eqref{eq:eqABb} to mix different perturbative orders,
\bqa
\label{eq:Lpert}
\Lr= \sum_{i= -1}^{\ell -1} X_i \alpha^i.
\eqa
Note that the loop and energy expansions should not be considered independently
and that the coefficients of $\Lr^k$ should be known up to the $(\ell +1)$ order. Equating the powers of $\alpha$ and $\lambda^\mj$ gives a system of equations to be fulfilled by the unknown coefficients $X_i$. As discussed in the previous section, only kinematics independent solutions $X^\prime_i$ for the $X_i$ are compatible with FDR. This determines the necessary and sufficient conditions for the matching of \eqref{eq:match}. For instance, when $\ell=1$,
\bqa
\label{eq:cond1}
&&
 \left\{
\begin{tabular}{l}
\hskip -5pt $A_{01}^\mjone-B_{111}^\mjone X^\prime_{-1}-B_{212}^\mjone {(X^\prime_{-1})}^2= 0$,  \\ \nonumber \\
\hskip -5pt $A_{11}^\mjone-B_{110}^\mjone-B_{111}^\mjone X^\prime_0-B_{211}^\mjone X^\prime_{-1}-2 B_{212}^\mjone X^\prime_{-1} X^\prime_0 = 0$,
\end{tabular} \right. \boldmath{ \forall \mjone} \nl
&& \nl
&& \hskip 10pt \frac{\partial X^\prime_{-1}}{\partial s_n}=
    \frac{\partial X^\prime_{0}}{\partial s_n} = 0.
\eqa
Since there are more equations than unknowns, relations must exist among
coefficients. If the conditions in \eqref{eq:cond1} are all obeyed, the solution is
\bqa
\ln\big({{\mur^\prime}^2}\big)=
\frac{X^\prime_{-1}}{\alpha}
+ X^\prime_{0}.
\eqa

Consider now a further independent amplitude $B_i$ computed in the effective model and evaluated at $\mur=\mur^\prime$, as in Fig. \ref{fig:3}. An interesting question is whether or not it reproduces the result of a calculation performed within the renormalizable theory, namely whether
\bqa
\label{eq:matchi}
B^{\rm \ell-loop}_{i}(\lambda,\alpha,\mur^\prime)= A^{\rm \ell-loop}_{i}(\lambda,\alpha).
\eqa
In \cite{Pittau:2019bba} a conjecture is formulated which states that
\eqref{eq:matchi} holds whenever the two amplitudes coincide at $\lambda= 0$,
\bqa
\label{eq:matchi0}
B^{\rm \ell-loop}_{i}(0,\alpha,\mur^\prime)= A^{\rm \ell-loop}_{i}(0,\alpha).
\eqa
In what follows, I describe a realistic model, in which this conjecture is verified, for a given class of loop corrections, to all perturbative orders.

\begin{figure}[t]
\vspace{-2.7cm}
\hskip -200pt
\includegraphics[width=1.9\textwidth, angle= {0}]{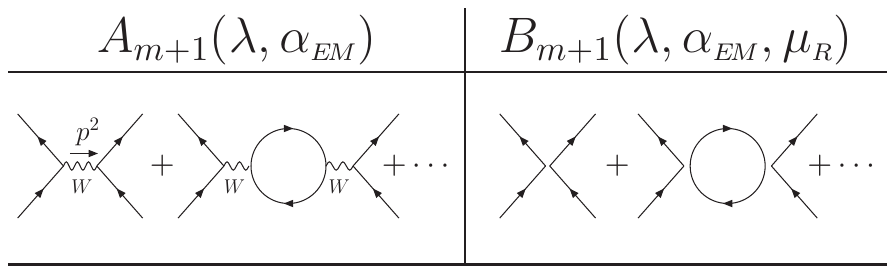}
\vspace{-15.7cm}
\caption{Example of diagrams contributing to charged current one-fermion-loop mediated interactions in the renormalizable and nonrenormalizable theories
  of \eqref{eq:Lexact} and \eqref{eq:Leff}. Comparing $A_{m+1}(\lambda,\alpha_{\scriptscriptstyle E\!M})$ with $B_{m+1}(\lambda,\alpha_{\scriptscriptstyle E\!M},{ \mur})$
  at $p^2 \ne 0$ allows one to find the exact solution of \eqref{eq:match}.} 
\label{fig:4}
\end{figure}

\begin{figure}[t]
\vspace{-6.1cm}
\hskip -95pt
\includegraphics[width=2.2\textwidth, angle= {0}]{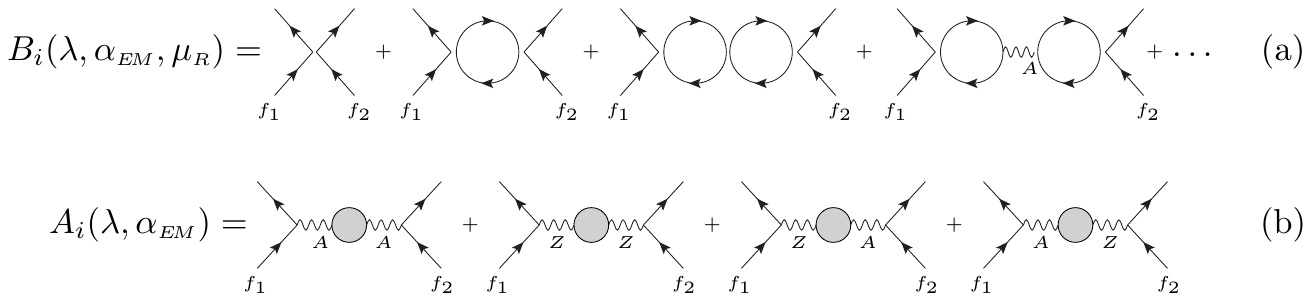}
\vspace{-14.2cm}
\caption{The neutral one-fermion-loop mediated amplitudes used to test the validity of \eqref{eq:matchi}. They are computed by using the Lagrangians of \eqref{eq:Leff} and \eqref{eq:Lexact}, respectively.} 
\label{fig:5}
\end{figure}

The effective Lagrangian
\bqa
\label{eq:Leff}
    {\cal L}^{\scriptstyle \rm N}(g^2,M^2,s^2_\theta)=
    {\cal L}^{\mbox{\tiny  QED }}+{ {\cal L}^{\mbox{\tiny FERMI  }}},~~~
{\cal L}^{\mbox{\tiny  FERMI}}=
  -\frac{g^2}{8 M^2} {J^\dag}_{\!\!\!c\alpha} J_{c}^\alpha -\frac{g^2}{8 M^2} J_{n\alpha} J_{n}^\alpha,
\eqa
is used to compute resummed one-fermion-loop interactions between two massless fermions and the results are compared with the predictions of the complete standard model (SM). Thus, the renormalizable theory to be matched onto
${\cal L}^{\scriptstyle \rm N}(g^2,M^2,s^2_\theta)$ is
\bqa
\label{eq:Lexact}
    {\cal L}^{\scriptstyle \rm R}(g^2,M^2,s^2_\theta)= {\cal L}^{\scriptstyle \rm SM}.
\eqa
Both Lagrangians in \eqref{eq:Leff} and \eqref{eq:Lexact} depend on the same set of bare parameters $\{g^2,M^2,s^2_\theta\}$, which can be fixed
experimentally it terms of the fine structure constant $\aem$, measured in the Thomson limit of the Compton scattering, the muon decay constant $G_F$, extracted from the muon lifetime, and the ratio $R_{\scriptscriptstyle e \nu}$ between the total $e^- \nu_\mu$ and $e^- \bar \nu_\mu$ elastic cross sections at zero momentum transfer. After that, one determines $\mur^\prime$ by computing any high-energy amplitude in which two massless fermions interact via charged fermion-loops, as in Fig. \ref{fig:4}.
The conditions in \eqref{eq:cond1} are fulfilled at the first order in $\lambda = p^2/ \hat M^2\ne 0$ when choosing
\bqa
\ln\big({{\mur^\prime}^2}\big)=
\frac{\pi {\hat s_\theta}^2}{\alpha_{\scriptscriptstyle E\!M}}
+ K_1,
\eqa
where ${\hat s_\theta}^2$, $\hat M^2$ are the tree-level solution for 
$s^2_\theta$, $M^2$ and
\bqa
\label{eq:K1}
K_1:= \frac{1}{2}
+\frac{\ln m^2_e +\ln m^2_\mu +\ln m^2_\tau}{12}
+\frac{\ln m^2_u +\ln m^2_c +\ln m^2_t}{6} 
+\frac{\ln m^2_d +\ln m^2_s +\ln m^2_b}{12}.
\eqa
As a matter of fact, in the case at hand $\Lr= \ln\big({{\mur^\prime}^2}\big)$ solves \eqref{eq:match} at any value of $\ell$ and $p^2$, if the one-fermion-loop contributions are resummed.
The independent amplitudes of \eqref{eq:matchi} are taken to be the
neutral current interactions between two arbitrary massless fermions $f_{1,2}$ illustrated in Fig. \ref{fig:5}.
They obey \eqref{eq:matchi0} and, when choosing $\mur= \mur^\prime$, 
the amplitude of Fig. \ref{fig:5}(a) coincides with the one of Fig. \ref{fig:5}(b) to all orders and energies, corroborating the conjecture of \cite{Pittau:2019bba}. 
Note that this is not trivial because \eqref{eq:matchi} is obeyed for any choice of $f_1$ and $f_2$, while the process of Fig. \ref{fig:4} only
involves $V-A$ interactions.

Finally, it is interesting to take advantage of the fact that an exact gauge invariant all-order solution for the bare parameters can be found by simply resumming one-fermion-loop diagrams. One finds, in particular \cite{Pittau:2019bba},
\bqa
s^2_\theta(\mur) =  {\hat s_\theta}^2\, \frac{F_1}{F_2},
\eqa
where
\bqa
F_1 :=  1 -\frac{\aem}{\pi {\hat s_\theta}^2} \left(\Lr-K_1 \right),~~~
F_2 :=  1 -\frac{8\aem}{3 \pi} \left(\Lr-K_2 \right),
\eqa
with $K_1$ in \eqref{eq:K1} and
\bqa
K_2 := \frac{1}{2}
+\frac{\ln m^2_e +\ln m^2_\mu +\ln m^2_\tau}{8}  
+\frac{\ln m^2_u +\ln m^2_c +\ln m^2_t}{6}  
+\frac{\ln m^2_d +\ln m^2_s +\ln m^2_b}{24}.
\eqa
Thus, $s^2_\theta(\mur^\prime)= 0$, namely the solution that matches
${\cal L}^{\scriptstyle \rm R}(g^2,M^2,s^2_\theta)$
onto ${\cal L}^{\scriptstyle \rm N}(g^2,M^2,s^2_\theta)$
is such that the number of parameters of the two theories is effectively reduced from three to two.

\section{Conclusions}
FDR can be used to perform higher-order calculations in renormalizable pQFTs.
NNLO results have been successfully reproduced for observables involving intermediate UV and IR divergences without going away from the physical four-dimensional space-time.

Under certain circumstances, loop corrections computed in high-energy renormalizable pQFTs can be matched onto low-energy nonrenormalizable Lagrangians ${\cal L}^{\scriptstyle \rm N}$ without modifying ${\cal L}^{\scriptstyle \rm N}$. This is possible only if UV infinities are handled {\em \`a la} FDR. For instance, ${\cal L}^{\scriptstyle \rm N}={\cal L}^{\mbox{\tiny  QED }}+{{\cal L}^{\mbox{\tiny FERMI}}}$ can be used to reproduce the exact electroweak interactions between two massless fermion lines induced by one-fermion-loop resummed gauge boson propagators. This is the first-ever example of nonrenormalizable pQFT consistently made predictive to all loop orders and energies without the addition of higher-dimensional operators.


\begin{thebibliography}{99}
\bibitem{Bollini:1972ui}
  C.~G.~Bollini and J.~J.~Giambiagi,
  \emph{Dimensional Renormalization: The Number of Dimensions as a Regularizing Parameter}, \href{https://link.springer.com/content/pdf/10.1007\%2FBF02895558.pdf}{\emph{Nuovo Cim.\ B}  {\bf 12} (1972) 20}.
%
\bibitem{tHooft:1972tcz}
  G.~'t Hooft and M.~J.~G.~Veltman,
  \emph{Regularization and Renormalization of Gauge Fields},
  \href{https://doi.org/10.1016/0550-3213(72)90279-9}{\emph{Nucl.\ Phys.\ B} {\bf 44} (1972) 189}.  
%
\bibitem{Cherchiglia:2010yd}
  A.~L.~Cherchiglia, M.~Sampaio and M.~C.~Nemes,
  \emph{Systematic Implementation of Implicit Regularization for Multi-Loop Feynman Diagrams},
  \href{https://doi.org/10.1142/S0217751X11053419}{\emph{Int.\ J.\ Mod.\ Phys.\ A} {\bf 26} (2011) 2591}
  [\href{http://arxiv.org/abs/arXiv:1008.1377}{\tt arXiv:1008.1377}].
%
\bibitem{Pittau:2012zd}
  R.~Pittau,
  \emph{A four-dimensional approach to quantum field theories},
  \href{https://doi.org/10.1007/JHEP11(2012)151}{\emph{JHEP} {\bf 1211} (2012) 151} [\href{https://arxiv.org/pdf/1208.5457.pdf}{\tt 1208.5457}].
%
\bibitem{Sborlini:2016gbr}
  G.~F.~R.~Sborlini, F.~Driencourt-Mangin, R.~Hernandez-Pinto and G.~Rodrigo,
  \emph{Four-dimensional unsubtraction from the loop-tree duality},
  \href{https://doi.org/10.1007/JHEP08(2016)160}{\emph{JHEP} {\bf 1608} (2016) 160} [\href{http://arxiv.org/abs/arXiv:1604.06699}{\tt 1604.06699}].
%
\bibitem{Gnendiger:2017pys}
  C.~Gnendiger {\it et al.},
  \emph{To ${d}$, or not to ${d}$: recent developments and comparisons of regularization schemes},
  \href{https://doi.org/10.1140/epjc/s10052-017-5023-2}{\emph{Eur.\ Phys.\ J.\ C} {\bf 77} (2017) no.7, 471}
  [\href{http://arxiv.org/abs/arXiv:1705.01827}{\tt 1705.01827}].
%
\bibitem{Cherchiglia:2012zp}
  A.~L.~Cherchiglia, L.~A.~Cabral, M.~C.~Nemes and M.~Sampaio,
  \emph{(Un)determined finite regularization dependent quantum corrections: the Higgs boson decay into two photons and the two photon scattering examples},
  \href{https://doi.org/10.1103/PhysRevD.87.065011}{\emph{Phys.\ Rev.\ D} {\bf 87} (2013) no.6,  065011}
  [\href{http://arxiv.org/abs/arXiv:1210.6164}{\tt 1210.6164}].
%
\bibitem{Donati:2013iya}
  A.~M.~Donati and R.~Pittau,
  \emph{Gauge invariance at work in FDR: $H \to \gamma \gamma$},
  \href{https://doi.org/10.1007/JHEP04(2013)167}{\emph{JHEP} {\bf 1304} (2013) 167} [\href{http://arxiv.org/abs/arXiv:1302.5668}{\tt 1302.5668}].
%
\bibitem{Pittau:2013qla}
  R.~Pittau,
  \emph{QCD corrections to $H \to gg$ in FDR},
  \href{https://doi.org/10.1140/epjc/s10052-013-2686-1}{\emph{Eur.\ Phys.\ J.\ C} {\bf 74} (2014) no.1,  2686}
  [\href{http://arxiv.org/abs/arXiv:1307.0705}{\tt 1307.0705}].
%
\bibitem{Donati:2013voa}
  A.~M.~Donati and R.~Pittau,
  \emph{FDR, an easier way to NNLO calculations: a two-loop case study},
  \href{https://doi.org/10.1140/epjc/s10052-014-2864-9}{\emph{Eur.\ Phys.\ J.\ C} {\bf 74} (2014) 2864}
  [\href{http://arxiv.org/abs/arXiv:1311.3551}{\tt 1311.3551}].
%
\bibitem{Page:2015zca}
  B.~Page and R.~Pittau,
  \emph{Two-loop off-shell QCD amplitudes in FDR},
  \href{https://doi.org/10.1007/JHEP11(2015)183}{\emph{JHEP} {\bf 1511} (2015) 183} [\href{http://arxiv.org/abs/arXiv:1506.09093}{\tt 1506.09093}].  
%
\bibitem{Page:2018ljf}
  B.~Page and R.~Pittau,
  \emph{NNLO final-state quark-pair corrections in four dimensions},
  \href{https://doi.org/10.1140/epjc/s10052-019-6865-6}{\emph{Eur.\ Phys.\ J.\ C} {\bf 79} (2019) no.4,  361}
  [\href{http://arxiv.org/abs/arXiv:1810.00234}{\tt 1810.00234}].
%
\bibitem{Driencourt-Mangin:2017gop}
  F.~Driencourt-Mangin, G.~Rodrigo and G.~F.~R.~Sborlini,
  \emph{Universal dual amplitudes and asymptotic expansions for $gg\rightarrow H$ and $H\rightarrow \gamma \gamma $ in four dimensions},
  \href{https://doi.org/10.1140/epjc/s10052-018-5692-5}{\emph{Eur.\ Phys.\ J.\ C} {\bf 78} (2018) no.3,  231}
  [\href{http://arxiv.org/abs/arXiv:1702.07581}{\tt 1702.07581}].
%
\bibitem{Driencourt-Mangin:2019aix}
  F.~Driencourt-Mangin, G.~Rodrigo, G.~F.~R.~Sborlini and W.~J.~Torres Bobadilla,\emph{Universal four-dimensional representation of $H \to \gamma \gamma$ at two loops through the Loop-Tree Duality},
  \href{https://doi.org/10.1007/JHEP02(2019)143}{\emph{JHEP} {\bf 1902} (2019) 143}
  [\href{http://arxiv.org/abs/arXiv:1901.09853}{\tt 1901.09853}].
%
\bibitem{Pittau:2013ica}
  R.~Pittau,
  \emph{On the predictivity of the non-renormalizable quantum field theories},
  \href{https://doi.org/10.1002/prop.201400079}{\emph{Fortsch.\ Phys.}  {\bf 63} (2015) 132}
  [\href{http://arxiv.org/abs/arXiv:1305.0419}{\tt 1305.0419}].
%
\bibitem{Pittau:2019bba}
  R.~Pittau,
  \emph{Matching high-energy electroweak fermion loops onto the Fermi theory without higher dimensional operators},
  \href{https://doi.org/10.1016/j.nuclphysb.2019.114835}{\emph{Nucl.\ Phys.\ B} {\bf 950} (2020) 114835}
  [\href{http://arxiv.org/abs/arXiv:1902.01767}{\tt 1902.01767}].
\end{thebibliography}
\end{document}